\documentclass[twocolumn]{autart}    

\usepackage{graphicx}          
                               
\usepackage{amsmath,amsfonts,amssymb,bbold}
\usepackage{algorithm}
\usepackage[noend]{algpseudocode}
\usepackage[square,sort,comma,numbers]{natbib}
\begin{document}

\begin{frontmatter}
	\runtitle{Cepstrum Distance for Deterministic Inputs}  
	
	\title{Applicability and interpretation of the deterministic weighted cepstral distance 
	} 
	
	
	\author[Leuven,imec]{Oliver Lauwers}\ead{oliver.lauwers@esat.kuleuven.be},    
	\author[Leuven,imec]{Bart De Moor}\ead{bart.demoor@esat.kuleuven.be},               
	
	\address[Leuven]{KU Leuven, Department of Electrical Engineering (ESAT), STADIUS Center for Dynamical Systems, Signal Processing and Data Analytics, Kasteelpark Arenberg 10, Leuven, Belgium}
	\address[imec]{imec, Leuven, Belgium}   


\begin{abstract}                          
Quantifying similarity between data objects is an important part of modern data science. Deciding what similarity measure to use is very application dependent. In this paper, we combine insights from systems theory and machine learning, and investigate the weighted cepstral distance, which was previously defined for signals coming from ARMA models. 

We provide an extension of this distance to invertible deterministic linear time invariant single input single output models, and assess its applicability. We show that it can always be interpreted in terms of the poles and zeros of the underlying model, and that, in the case of stable, minimum-phase, or unstable, maximum-phase models, a geometrical interpretation in terms of subspace angles can be given. We then devise a method to assess stability and phase-type of the generating models, using only input/output signal information. In this way, we prove a connection between the extended weighted cepstral distance and a weighted cepstral model norm.

In this way, we provide a purely data-driven way to assess different underlying dynamics of input/output signal pairs, without the need for any system identification step. This can be useful in machine learning tasks such as time series clustering. An iPython tutorial is published complementary to this paper, containing implementations of the various methods and algorithms presented here, as well as some numerical illustrations of the equivalences proven here.
\end{abstract}
\end{frontmatter}
\section{Introduction}

Quantifying similarity between two data objects is a quintessential part of system identification, control theory and machine learning. Whether it is to assess how good an estimated system fits a dataset, decide whether the output of a system is close enough to the reference signal, or to discern relations between different data objects, always, an implicit or explicit choice of similarity measure has to be made.

For signals coming from dynamical systems, which are traditionally handled with techniques from the field of systems and control, it is important to keep track of the dynamics of the underlying generative systems. Indeed, in systems and control applications, it are precisely these dynamics that interest us, as they give us information about normal operation characteristics, changes in system behaviour and many other system properties.

However, in modern machine learning applications, these insights about the correlation in time by which time series are characterized, are rarely taken into account. Rather, signals - or time series, these terms will be used interchangeably in this paper - are either treated as ordinary vectors, and a vector distance (e.g. the Euclidean distance) is employed, or one or more features or statistics of the signals are calculated (e.g. mean, median, standard deviation), and a distance measure on these features is employed, largely ignoring the dynamics of the generative models. On the other hand, explicitly identifying systems is infeasible and computationally expensive in typical machine learning applications, characterized by large amounts of very long time series, often in need of an automated solution, without taking into account much domain knowledge.

In this paper, we argue that there is a need for dynamics-based distance measures in machine learning applications in general and time series clustering specifically. We then look at one such measure, the weighted cepstral measure \cite{boets2007mutual,boets2005clustering,DeCockThesis,de2002subspace,martin2000metric}, which was previously and up to now only defined for signals generated by ARMA models, driven by white noise inputs. It is used in various clustering and classification tasks, such as human activity recogniton \cite{Chaudhry_2013_CVPR_Workshops,liu2016extreme}, dynamic texture recognition \cite{saisan2001dynamic,Yang20161310} and structural damage identification \cite{STC:STC235}.

The main contributions of this paper are:
\begin{itemize}
	\item We extend the weighted cepstral distance to general invertible deterministic linear time invariant (LTI) single input single output (SISO) systems.
	\item We extend the distance measure to signals coming from unstable, maximum-phase systems.
	\item We prove that these extensions can still be interpreted geometrically with the notion of \emph{subspace angles}, and system theoretically in terms of poles and zeros of the generative models. This links the weighted cepstral distance to the weighted cepstral norm.
	\item We interpret the distance in terms of poles and zeros for systems with a mixture of stable and unstable poles and/or minimum and maximum-phase zeros. The interpretation in terms of subspace angles is lacking in this case.
	\item We provide a purely data-driven way to assess whether a signal comes from a completely minimum-phase/stable, a completely maximum-phase/unstable, or a mixed model, by employing what is known as the \emph{complex cepstrum}.
\end{itemize}

This paper is organised as follows. In Section \ref{sec:clustering}, we discuss an example of a machine learning task that is critically dependent on by the distance measure considered: clustering time series. In Section \ref{sec:notationdefinitions}, we introduce some notation and define some notions we will use throughout. Section \ref{sec:motivatingexample} gives a motivating example of how taking into account underlying dynamics can be critical in a clustering problem. This then leads us to consider a model norm, the weighted cepstral distance, in Section \ref{sec:cepstrum}, connecting it to a time series distance at the end of that Section. We explain why we need to assess whether systems are minimum, maximum or mixed-phase. Section \ref{sec:phasetype} introduces a novel data-driven approach to asses the phasetype of the underlying model based on input/output signals. The paper is accompanied by an iPython notebook tutorial\footnote{https://github.com/Olauwers/Applicability-and-interpretation-of-the-deterministic-weighted-cepstral-distance}, results of which are briefly discussed in Section \ref{sec:numerical}. Conclusions and future paths of research can be found in Section \ref{sec:conclusions}. The Appendices contain some computational and theoretical details of the implementation of the different methods, and the proofs of the several equivalences.

\section{Clustering signals}
\label{sec:clustering}

Clustering signals is the (unsupervised) task of finding groups of similar time series in a dataset, and is an important topic in contemporary machine learning. Traditional clustering methods for other types of data do not carry over trivially, as signal datasets typically are high-dimensional, and temporal correlations both between signals and within signals need to be taken into account when clustering.

Time series clustering algorithms consist of three main components: 
\begin{itemize}
\item a similarity measure based on relevant features of the data, 
\item a clustering scheme, 
\item a way to evaluate the clustering results.
\end{itemize}
While the latter two might carry over from other data types, or can be generic across applications, the notion of similarity depends critically on the specific problem at hand.

From the point of view of systems and control theory, we often are interested in the dynamics underlying the signal, and not so much in the specific shape of the signal. If we had access to the generating model, we could calculate distances based on model norms, such as the $H_2$ or $H_\infty$ norms, which explicitly takes into account the model dynamics. However, when only input and output signals are accessible, we resort to distance measures that can be calculated from the raw data alone, or explicitly identify a model of the system, which can be computationally expensive and typically requires quite some user expertise and intervention (some norms, such as the $H_2$ norm and the weighted cepstral norm, can be calculated from input/output signals, and as such do not require this step). In practice, the problem is often not given much attention and off-the-shelf distance measures that are sometimes ill-suited to handle the application at hand are used, such as shape-based distances, like the Euclidean distance (see \citep{pereira2013common} for a discussion).

To avoid the latter, we are thus interested in distances based on model norms that can be calculated from raw data, in an automated way (without user chosen design parameters, model order estimation, ...). One such distance is the weighted cepstral distance \citep{boets2007mutual,boets2005clustering,DeCockThesis,de2002subspace,martin2000metric}, which was proven to be a model norm in the case of data generated by ARMA models. We will formally introduce the data type we handle in Section \ref{sec:notationdefinitions}, as well as the weighted cepstral distance. We proceed to illustrate why it is useful in systems and control theory with a motivating example in Section \ref{sec:motivatingexample}.

\section{Notation and definitions}
\label{sec:notationdefinitions}

In this Section, we first introduce some general definitions and notation, then discuss the cepstrum. Afterwards, we introduce the notion of subspace angles, and we end the Section by introducing some distance measures.

\subsection{General notation}
Define $y$ to be an \emph{output signal}, which is the output of a system driven by the \emph{input signal} $u$. $y(k)$ is then the value of the output signal at \emph{timepoint} $k$, and similarly for $u(k)$. Both signals $y$ and $u$ are of length $N$. The output signal is generated by a \emph{Linear Time Invariant (LTI) Single-Input Single-Output (SISO) dynamical system}, of which the dynamics can be described by the \emph{state-space model}
\begin{equation}
\left\{
\begin{aligned}
x(k+1) &= Ax(k) + Bu(k)\\
y(k) &= Cx(k) + Du(k)
\label{eq:timedomain}
\end{aligned}
\right.
,
\end{equation}
with $x(k) \in \mathbb{R}^n$ the \emph{states} of the model and $A$, $B$, $C$ and $D$ \emph{system matrices} of appropriate dimensions.
We assume this model is invertible. Furthermore, we define for every state-space model an observability matrix $\Gamma_j$ as
\begin{equation}
\Gamma_j = \left(\begin{matrix}
C & CA & CA^2 & \cdots & CA^{j-1}
\end{matrix}\right)^\intercal,
\label{eq:observability}
\end{equation}
where the $A$ and $C$ matrix are the matrices from Equation \eqref{eq:timedomain}, and $\cdot^\intercal$ denotes the transpose.

Taking the \emph{z-transform} of the signal, the relation between the input and output becomes, assuming $x(0) = 0$, (in \emph{frequency domain}, denoted with the variable $z$)
\begin{equation}
Y(z) = H(z)U(z),
\label{eq:frequencydomain}
\end{equation}
with $H(z)$ the \emph{transfer function} of the system, written in terms of system matrices as 
\begin{equation}
H(z) = D + C\left(z\mathbb{1} - A\right)^{-1}B.
\end{equation}
For a SISO LTI system, the transfer function is a rational function in $z$, with both numerator and denominator a polynomial. We can express such a system as
\begin{equation}
Y(z) = \frac{b(z)d(z)}{a(z)c(z)}U(z).
\end{equation}
Here, the polynomial $b(z)$ has roots of a magnitude smaller than 1 (called the \emph{minimum-phase zeros}, denoted by $\beta_j$), the polynomial $d(z)$ has roots of magnitude greater than 1 (\emph{maximum-phase zeros}, $\delta_j$). Similarly, the polynomial $a(z)$ in the denominator has roots of magnitude smaller than 1 (\emph{stable poles}, $\alpha_j$) and $c(z)$ has roots of magnitude greater than 1 (\emph{unstable poles}, $\gamma_j$).\footnote{We assume, throughout the paper, that there are no poles or zeros exactly on the unit circle (i.e. for $z = e^{i\omega}$, with $i$ the imaginary unit). Since, in the discrete frequency domain, we evaluate at the unit circle, not all notions introduced here will be defined for systems with poles or zeros on that circle. The cepstrum, for example, is not well defined there. These cases will therefore not be considered here. For a discussion on how to deal with these kinds of roots, see \citep{oppenheim1975digital}.}

Written out in terms of poles and zeros, the transfer function becomes
\begin{equation}
H(z) = g\frac{\prod_{j=1}^{q}\left(1 - \beta_j z^{-1}\right)\prod_{j=1}^{r}\left(1 - \delta_j z^{-1}\right)}{\prod_{j=1}^{p}\left(1 - \alpha_j z^{-1}\right)\prod_{j=1}^{s}\left(1 - \gamma_j z^{-1}\right)},
\label{eq:poleszeros}
\end{equation}
where $q$, $r$, $p$, $s$ denote the degree of the corresponding polynomial of the respective types of zeros and poles and $g$ is a constant factor called the \emph{gain} of the transfer function.

The \emph{power spectrum} of the system with transfer function $H$ is defined as
\begin{equation}
\Phi_h(\textnormal{e}^{i\omega}) = H(\textnormal{e}^{i\omega})\overline{H(\textnormal{e}^{i\omega})} = \left|H(\textnormal{e}^{i\omega})\right|^2,
\label{eq:powerspectrum}
\end{equation}
and similarly for $\Phi_y$, the power spectrum of the output, and $\Phi_u$, for the input. Here the overbar denotes the complex conjugate, $i$ the imaginary unit and $|\cdot|$ denotes magnitude. A property of the power spectrum is that $\Phi_h(z) \geq 0,\hspace{0.4em} \forall z \in e^{i\omega}$. Furthermore, we have from Equation \eqref{eq:frequencydomain} that
\begin{equation}
\Phi_y(\textnormal{e}^{i\omega}) = \Phi_h(\textnormal{e}^{i\omega})\Phi_u(\textnormal{e}^{i\omega})
\label{eq:spectrumdomain}
\end{equation}

The assumption of invertibility of the system amounts to assuming $D \neq 0$ in Equation \eqref{eq:timedomain}. The equivalent of Equation \eqref{eq:timedomain} for the inverse system is then given by
\begin{equation}
\left\{
\begin{aligned}
x(k+1) &= \left(A-BD^{-1}C\right)x(k) + BD^{-1}y(k)\\
u(k) &= -D^{-1}Cx(k) + D^{-1}y(k)
\label{eq:timedomaininverse}
\end{aligned}
\right.
.
\end{equation}
The corresponding transfer function of the inverse system can be written respectively in terms of system matrices and pole and zero polynomials as
\begin{equation}
\begin{aligned}
&H^{-1}(z)\\
&= D^{-1} - D^{-1}C(z\mathbb{1} - A +BD^{-1}C)^{-1}BD^{-1}\\
&=\frac{a(z)c(z)}{b(z)d(z)}.
\end{aligned}
\end{equation}
From these equations, we can extend all definitions in this section to the inverse system.

\subsection{Cepstrum}
Denoting the \emph{inverse Fourier transform} as $\mathcal{F}^{-1}$, the power cepstrum\footnote{The terminology power spectrum stems from the fact that it is based on the power spectrum. A different notion, called \emph{complex cepstrum}, will be introduced later on. We will use the terms \emph{power cepstrum} and \emph{cepstrum} interchangeably. When we refer to the complex cepstrum, we will always write it out explicitly.} $c_h$, of a transfer function $H(z)$ is written as
\begin{equation}
c_h(k) = \mathcal{F}^{-1}(\log\Phi_h(\textnormal{e}^{i\omega})),
\label{eq:definitionpowercepstrum}
\end{equation}
again adopting similar definitions for $c_y$ and $c_u$, the power cepstra of respectively output and input.

The rationale behind employing the power cepstrum in signal processing stems from a subfield called \emph{homomorphic signal processing} \citep{oppenheim1975digital}, where the objective is to simplify complicated multiplicative operators (such as convolutions). As we can see from Equation \eqref{eq:definitionpowercepstrum}, the convolution from time domain changes into a multiplication by virtue of the property in Equation \eqref{eq:spectrumdomain}. The logarithm takes this multiplication to an addition. Finally, the inverse Fourier transform is applied to convert the problem back to (a transformation of) time domain. This type of analysis is often referred to as \emph{quefrency alanysis} \citep{bogert1963quefrency}.

Starting from equation \eqref{eq:poleszeros}, we can express the power cepstrum $c_h(k)$ as (see Appendix \ref{app:derivepowerspectrum} for a derivation)
\begin{equation}
\begin{aligned}
c_h(k) = &\sum_{j=1}^{p}\frac{\alpha_j^{|k|}}{|k|} + \sum_{j=1}^{s}\frac{\gamma_j^{-|k|}}{|k|} \\
- &\sum_{j=1}^{q}\frac{\beta_j^{|k|}}{|k|} - \sum_{j=1}^{r}\frac{\delta_j^{-|k|}}{|k|}
\label{eq:powercepstrumcoefficients}
\end{aligned} \hspace{7pt} \forall k \neq 0,
\end{equation}
and
\begin{equation}
c_h(0) = g^\prime,
\label{eq:zerothcoefficient}
\end{equation}
which is a combination of the gain of equation \eqref{eq:poleszeros} and some rest terms coming from the maximum-phase zeros and unstable poles. This term is not too important for our purposes, and we will omit futher discussion. The interested reader is referred to \citep{oppenheim1975digital}. Note that the poles and zeros of magnitude greater than 1 appear as their inverse. To see why this is so, we refer to Appendix \ref{app:derivepowerspectrum}.

As an aside, we can (partly) give an interpretation of these cepstrum coefficients in terms of system matrices. Indeed, note that in the minimum-phase, stable case, the poles of a model are the eigenvalues of the $A$-matrix in state space representation (Equation \eqref{eq:timedomain}). The zeros are the poles of the inverse model (Equation \eqref{eq:timedomaininverse}), and therefore the eigenvalues of $A-BD^{-1}C$. All these eigenvalues are, of course, of magnitude smaller than 1.

We can then write, using only the $\alpha_j$'s and $\beta_j$'s in Equation \eqref{eq:powercepstrumcoefficients} (i.e., the breakpoints with magnitude smaller than 1),
\begin{equation}
c_h(k) = \textnormal{tr}\left\{\frac{A^k}{k}\right\} - \textnormal{tr}\left\{\frac{\left(A-BD^{-1}C\right)^k}{k}\right\},
\label{eq:cepstrumsystemmatrices}
\end{equation}
where $\textnormal{tr}\{\cdot\}$ denotes the trace of a matrix.

We now have a simple relation for the cepstrum coefficients:
\begin{equation}
c_h(k) = c_y(k) - c_u(k) \hspace{7pt} \forall k,
\end{equation}
which allows us to compute the cepstrum coefficients of the transfer functions from the input-output signal pair alone.  In principle, the power cepstrum can be calculated from input/output data and then related via \eqref{eq:powercepstrumcoefficients} to poles and zeros of the underlying LTI system. 
 
In analogy to the power cepstrum, we denote the \emph{complex cepstrum}\footnote{Note that the term complex cepstrum is a bit of a misnomer, as the complex cepstrum coefficients of a real signal are real. The name stems from the fact that in the complex cepstrum, information on the phase of the system is retained, which is not the case in the power cepstrum, as poles and zeros of magnitude greater than 1 appear as their inverses in Equation \eqref{eq:powercepstrumcoefficients}.} of a transfer function by $\hat{c}_h$, and write
\begin{equation}
\hat{c}_h = \mathcal{Z}^{-1}(\log H(z)),
\label{eq:definitioncomplexcepstrum}
\end{equation}
where $\mathcal{Z}^{-1}$ denotes the \emph{inverse z-transform}.

A similar derivation as the one for the power cepstrum results in
\begin{equation}
\hat{c}_h(k) = \left\{\begin{aligned}
\sum_{j=1}^{p}\frac{\alpha_j^k}{k} &- \sum_{j=1}^{q}\frac{\beta_j^k}{k} & \forall k>0 \\
\log(g^\prime)& & k=0 \\
- \sum_{j=1}^{s}\frac{\gamma_j^{k}}{k} &+ \sum_{j=1}^{r}\frac{\delta_j^{k}}{k} & \forall k<0 
\end{aligned}\right. .
\label{eq:complexcepstrumcoefficients}
\end{equation}

We provide a detailed description on how to numerically compute the power and complex cepstra in Appendix \ref{app:computecepstrum}.

\subsection{Subspace angles}
\begin{figure}
	\centering
	\includegraphics[width=\linewidth]{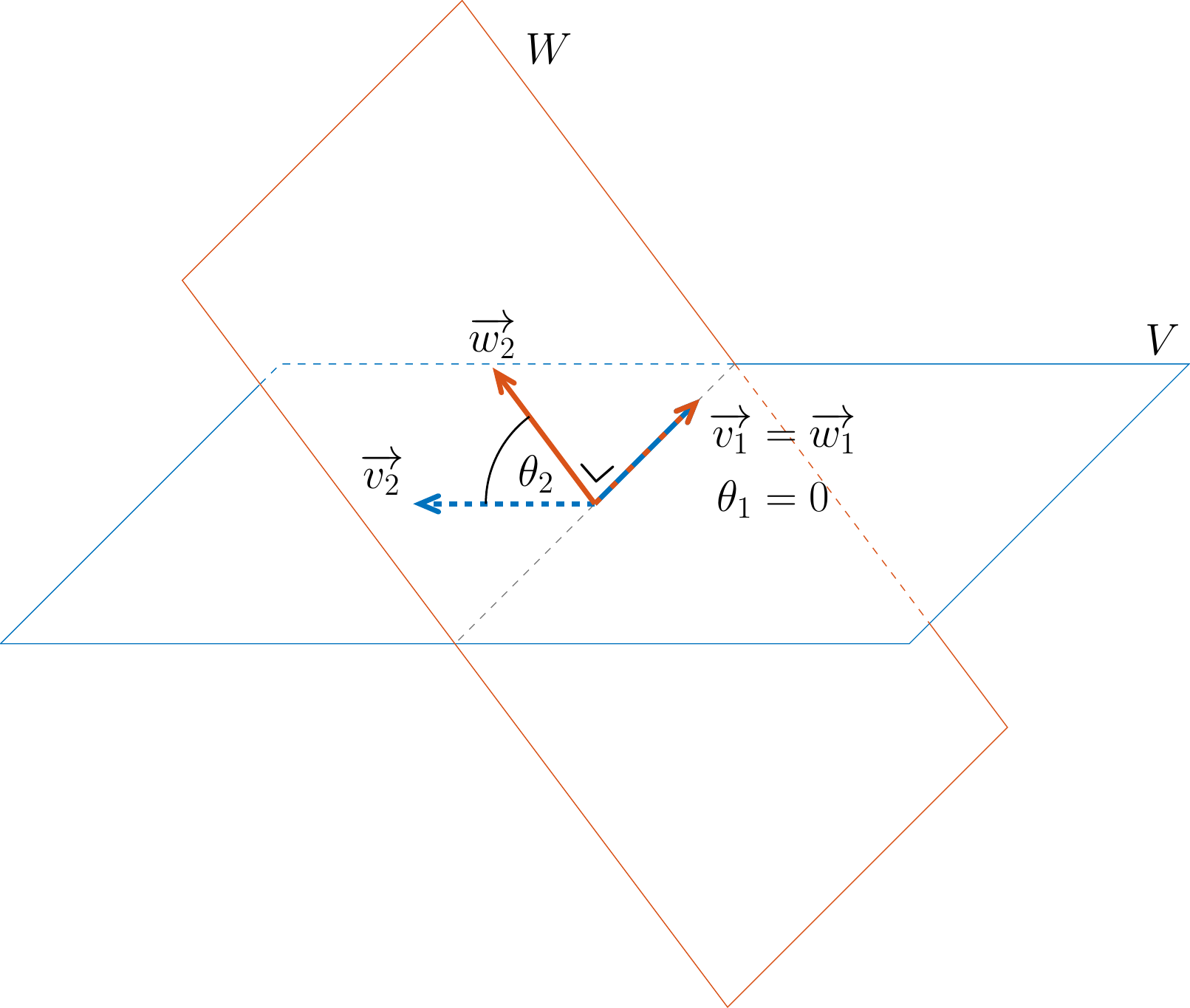} 
	\caption{Given two subspaces $V$ and $W$, we calculate the angles between them by first choosing unit vectors, called principal directions, $\protect\overrightarrow{v_1} \in V$ and $\protect\overrightarrow{w_1} \in W$, such that $\theta_1 = \angle(\protect\overrightarrow{v_1},\protect\overrightarrow{w_1})$, the angle between the two vectors, is minimised (in this case, it is 0). We then iteratively find new principal directions $\protect\overrightarrow{v_i} \in V$ and $\protect\overrightarrow{w_i} \in W$ in $V$ and $W$, perpendicular to all previous principal directions, such that the angle $\theta_i=\angle(\protect\overrightarrow{v_i},\protect\overrightarrow{w_i})$ between them is minimised. These angles $\theta_i$ are the principal angles between subspaces $V$ and $W$ and can be calculated in general from an eigenvalue problem.}
	\label{fig:subspaceangles}                                
\end{figure}

\emph{Principal angles} are generalizations of angles between vectors, and describe angles between given subspaces of a vector space, going back to Jordan \citep{jordan1875essai}, Hotelling \citep{hotelling1936relations} and Akaike \citep{akaike1974stochastic}. Figure \ref{fig:subspaceangles} shows a visualisation of these angles, along with a short description of how to obtain them. They are defined iteratively as follows.

Denote the principal angles between two subspaces, $V$ and $W$, by $\theta_k$, with $k$ the dimension of the smallest subspace. Take unit vectors, $\overrightarrow{v}\in V$ and $\overrightarrow{w}\in W$. We then define recursively the angles $\theta_k$ as
\begin{equation}
\begin{aligned}
\cos(\theta_k) = {}&\underset{\overrightarrow{v}\in V,\overrightarrow{w}\in W}{\text{maximize}}
&  \overrightarrow{v}^\intercal &\cdot \overrightarrow{w} \\
& \text{subject to}
&  ||\overrightarrow{v}|| &= ||\overrightarrow{w}|| = 1, \\
& & \overrightarrow{v}^\intercal &\cdot \overrightarrow{v_i} = 0 \hspace{1em}\forall i<k, \\
& & \overrightarrow{w}^\intercal &\cdot \overrightarrow{w_i} = 0 \hspace{1em}\forall i<k.
\end{aligned}\hspace{-0.5em}
\end{equation}
The $v_i$ and $w_i$ found this way are called \emph{principal directions}.

These principal angles give a measure of (dis)similarity between subspaces. The smaller the angles, the more similar or \emph{closer} the subspaces are. If all principle angles were right angles, all principal directions would be perpendicular to each other, and there would be no shared dimensions in the subspaces.

As shown in \citep{DeCockThesis,de2000subspace2,van2012subspace}, the squared cosines of the principal angles and the principal directions between the column spaces of two full rank matrices $A \in \mathbb{R}^{m\times }$ and $B \in \mathbb{R}^{m\times }$ can be calculated from the symmetric generalized eigenvalue problem
\begin{equation}
\begin{pmatrix}
0 & A^\intercal B\\
B^\intercal A & 0
\end{pmatrix}
\begin{pmatrix}
x\\
y
\end{pmatrix} = \lambda
\begin{pmatrix}
A^\intercal A & 0\\
0 & B^\intercal B
\end{pmatrix}
\begin{pmatrix}
x\\
y
\end{pmatrix},
\end{equation}
subject to $x^\intercal A^\intercal A x = 1$ and $y^\intercal B^\intercal B y = 1$.
The solutions $\lambda_i$ of this eigenvalue problem are the cosines of the principal angles $\theta_i$ between the column spaces of $A$ and $B$.

From here, one can show that
\begin{equation}
\cos^2\theta_i = \lambda_i\left((A^\intercal A)^{-1}A^\intercal B(B^\intercal B)^{-1}B^\intercal A\right),
\label{eq:subspaceeigenvalues}
\end{equation}
where $\lambda_i(\cdot)$ denotes the $i$-th eigenvalue of the expression between brackets.

The \emph{subspace angles} between two models, $M_1$ and $M_2$ are defined as the principal angles between the column spaces of the following infinite observability matrices:
\begin{equation}
\begin{aligned}
&\left[M_1 \sphericalangle M_2\right] =\\
&\left[\left(\Gamma_\infty\left(M_1\right)\Gamma_\infty\left(M_2^{-1}\right)\right) \sphericalangle \left(\Gamma_\infty\left(M_2\right)\Gamma_\infty\left(M_1^{-1}\right)\right)\right]
\label{eq:anglesreformulated}\hspace{-0.5em}
\end{aligned}
\end{equation}
where $M^{-1}$ denotes the inverse of model $M$.

\subsection{Distance measures}
We now formally introduce two distance measures:
\begin{description}
	\item[\emph{Euclidean distance}] The Euclidean distance is based on the well-known $l_2$-norm, and given by 
	\begin{equation}
	d_e(y_1,y_2) = \sqrt{\sum_{k = 0}^{N}\left(y_1(k)-y_2(k)\right)^2}.
	\end{equation}
	This distance is purely based on the shape of the particular signal, not on its underlying dynamics, and thus not appropriate for solving problems were these dynamics are the most important factor.\footnote{Note that there is no clear way to take into account the inputs in this distance. Naive approaches like including a term with the euclidean distance between input signals did not change the conclusions presented in this paper. We therefore omit this matter in the rest of our discussion.} This is not a criticism against the Euclidean distance, only a statement about its application domain.
	\item[\emph{Cosine similarity}] The cosine similarity gives the cosine of the angle, $\theta$, between two vectors, i.e.
	\begin{equation}
	\begin{aligned}
	d_\theta(y_1,y_2) &= \cos(\theta)\\
	&= \frac{\sum_{k=0}^N y_1(k)y_2(k)}{\sqrt{\sum_{k = 0}^{N}y_1^2(k)}\sqrt{\sum_{k = 0}^{N}y_2^2(k)}}.
	\end{aligned}
	\end{equation}
	This similarity interprets the time series $y_1$ and $y_2$ as vectors, and calculates the angle between them. It then returns the cosine of this angle. Note that this is not, strictly speaking, a distance measure, as two orthogonal vectors, which are far apart, will return a right angle, and a cosine of 0. A proper distance would only return a value of 0 for identical objects. It is, however, a \emph{similarity} measure, and one could define a distance by transforming the end result. We will, however, not do so here, but rather continue with the similarity measure with the understanding that $d_\theta = 0$ means the two objects are \emph{far away} from each other, and $d_\theta = 1$ means two objects are identical to each other (as $\theta = 0$ in this case).
	\item[\emph{Weighted power cepstral distance}] The power cepstral distance is defined in terms of the power cepstrum coefficients of two signals as
	\begin{equation}
	d_c(y_1,y_2)^2 = \sum_{k=1}^{\infty}k\left(c_{h_1}(k) - c_{h_2}(k)\right)^2.
	\label{eq:cepstraldistance}
	\end{equation}
	From Equation \eqref{eq:powercepstrumcoefficients}, we can see this distance will be related to poles and zeros of the underlying models. This is an indication that the power cepstral distance measures the underlying dynamics of the systems. Indeed, in subsequent Sections, we will show that the power cepstral distance constitutes a model norm and provide a geometrical interpretation for this norm. 
	As can be readily seen from the definition, the zeroth cepstrum coefficients, $c_{h_1}(0)$ and $c_{h_2}(0)$ do not contribute to the distance, which is why we can omit a more detailed discussion of this zeroth term.
\end{description}

\section{Motivating example}
\label{sec:motivatingexample}

\begin{figure}[h]
		\centering
		\hspace{-0.5em}
		\includegraphics[width=\linewidth]{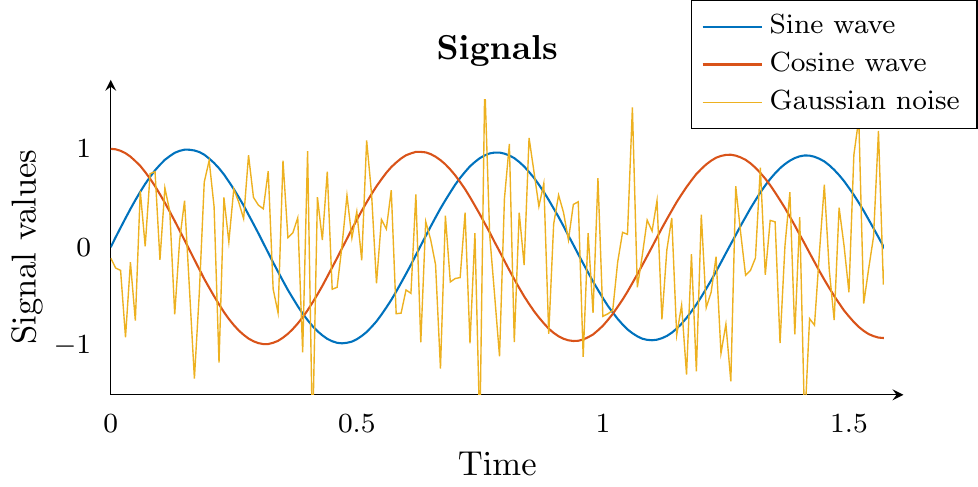} 
		\caption[]{Three signals that constitute the motivating example: exponentially damped gaussian noise, an exponentially damped\footnotemark sine wave at 10 rad/s and an exponentially damped cosine wave at 10 rad/s.}
		\label{fig:motivatingexample}                                
\end{figure}
\footnotetext{The damping is the result of the technical consideration that, analytically, the theoretical framework we develop here does not hold for poles and zeros on the unit circle. A slight damping is enough to alleviate this problem.}

\begin{table}
	\centering
	\caption{Euclidean, cosine and weighted cepstral distances between the different signals in Figure \ref{fig:motivatingexample}. The sine wave is denoted in the table by sin, the cosine by cos and the Gaussian white noise by Gauss. $d_e$ denotes the Euclidean distance, $d_\theta$ denotes the cosine similarity, while $d_c$ denotes the weighted cepstral distance, as defined in Section \ref{sec:notationdefinitions}. The signals were generated from $t=0$ to $t=11$, in increments of $0.01$. The white noise has a mean of $0$ and a standard deviation $\sigma = 0.7071$, and was generated in Matlab with the seed of the random number generator set to 1.}
	\vspace{0.5em}
	\label{tab:distances}
	\begin{tabular}{r|ccc}
		& d(sin,cos) & d(sin,Gauss) & d(cos,Gauss) \\ \hline
		$d_e$ & $664.2$ & $683.9$ & $702.2$ \\
		$d_\theta$ & $4.9\cdot10^{-3}$ & $-1.5\cdot10^{-3}$ & $-2.8\cdot10^{-2}$ \\
		$d_c$ & $1.9$ & $116.1$ & $116.7$ \\ \hline
	\end{tabular}
\end{table}

\begin{table}
	\centering
	\caption{Zeroth, first and second order statistics of the different signals. These statistics of the sine wave, the cosine wave and the random (White) signal, are all very close to each other, making them unsuitable features to devise a distance measure on.}
	\vspace{0.5em}
	\label{tab:statistics}
	\begin{tabular}{c|ccc}
		& Median & Mean & Standard deviation \\ \hline
		Sine  & $3.2\cdot10^{-2}$ & $1.4\cdot10^{-2}$ & $5.5\cdot10^{-1}$ \\
		Cosine  & $2.6\cdot10^{-3}$ & $6.1 \cdot 10^{-4}$ & $5.5\cdot10^{-1}$ \\ 
		White & $2.2\cdot10^{-3}$ & $-1.4 \cdot 10^{-2}$  & $5.5\cdot10^{-1}$ \\ \hline
	\end{tabular}
\end{table}

When clustering time series from a systems and control perspective, it is necessary to define a notion of similarity that takes into account information about the dynamics of the system. In this section, we present a simple motivating example that shows the need for a model-driven distance measure in systems and control theory.

Suppose we have an exponentially damped sine wave, with a pulsation of 10 rad/s. In our simple example, we have two more signals: an exponentially damped cosine wave, also with a pulsation of 10 rad/s, and an exponentially damped sequence of random numbers. Figure \ref{fig:motivatingexample} gives part of these signals. Note that there is no input signal provided for these signals, so they are autonomous systems (i.e., the input is a zero signal). A distance measure can then decide which of the latter two signals is most similar to the original sine wave. From a systems and control point of view, it is obvious this would be the cosine wave, as it differs from the original system only in initial state, but not in dynamics.\footnote{It is important to note that choosing the right distance measure is very much application-dependent. Other applications might rightly deem the random number sequence to be the most similar one. There is thus not one \emph{best} notion of distance or similarity, but rather a scale of measures varying in appropriateness for the problem at hand.}

As can be seen from Table \ref{tab:distances}, the Euclidean distance deems the similarity between all pairs of signals to be about the same. This is, of course, to be expected, as all vector pairs constitute orthogonal vector pairs. This is not a criticism of the Euclidean distance, but an argument against using it for problems where the underlying dynamics of the signals are critical.

The cosine distance between all signals is about 0. This, again, was to be expected, as sine and cosine constitute orthogonal vectors, and white noise signals are orthogonal to all others.

The weighted cepstral distance, however, correctly identifies the sine and cosine wave as coming from the same model, resulting in a distance close to zero (which would be the analytically expected value; the deviation is the result of the finite length of the series). Furthermore, it recognizes that the sine wave and Gaussian noise are far apart from a dynamical perspective. Interestingly, it returns more or less the same distance for both the sine and cosine wave compared to Gaussian noise, indicating again that it correctly quantifies the difference in dynamics between the signals.

From Table \ref{tab:statistics}, we see that the zeroth, first and second order statistics also do not provide a solution to firmly discern which signal pairs are closest. Again, for all signals, the results are about the same.

Of course, the example here is a toy problem, which can be easily solved by a trivial extension of the Euclidean distance (i.e. time-shifting both signals to minimize the distance between them). However, it does indicate that in systems and control theory, the choice of distance measure should be made with care. A more involved simulation example by the authors can be found in \citep{lauwers2017}. If the link between the measure and the underlying dynamics is not clear, the validity and usefulness of the clustering results may be in jeopardy. 

In what follows, we will interpret the weighted cepstral distance as a model norm in the case of SISO LTI models with known inputs. We then relate this model to the notion of \emph{subspace angles} between the models. We prove that the distance is a model norm in both the case of minimum-phase, stable and that of maximum-phase, unstable systems. We show some of the problems that arise in the mixed-phase case and a data-driven test to check the phase-type of the underlying dynamics.

\section{Cepstral norm}
\label{sec:cepstrum}

In this Section, we consider the power cepstral distance as a model norm, based on the distance measure in Equation \eqref{eq:cepstraldistance},
\begin{equation}
||H||_c^2 = \sum_{k=1}^{\infty}k\left(c_{h}(k)\right)^2.
\label{eq:cepstralnorm}
\end{equation}

Note that this is not, in fact, a proper model norm. Indeed, it does not take into account information in the zeroth cepstrum coefficient, i.e., information on scale, and only takes into account the power spectrum, which by virtue of Equation \eqref{eq:powerspectrum}, only takes into account the magnitude of the transfer function, and not it's angular components (which we call \emph{phase information}). However, it is a norm on the vector space constituted by equivalence classes containing all systems that have the same power spectrum \eqref{eq:powerspectrum}, up to a difference in gain.

As shown in \cite{DECOCK2003439}, the weighted cepstral norm can be linked to the Hilber-Schmidt-Hankel norm of the (double infinite) Hankel matrix of cepstrum coefficients
\begin{equation}
C_h = \begin{pmatrix}
c_h(1) & c_h(2) & c_h(3) & \ldots \\
c_h(2) & c_h(3) & c_h(4) & \ldots \\
c_h(3) & c_h(4) & c_h(5) & \ldots \\
\vdots & \vdots & \vdots & \ddots
\end{pmatrix} \in \mathbb{R}^{\infty \times \infty}.
\end{equation}
The Hilbert-Schmidt norm of $C_h$ is
\begin{equation}
||C_h||^2_{HS} = \textnormal{tr}C_hC_h^\intercal = \sum_{k=1}^{\infty}k\left(c_{h}(k)\right)^2,
\end{equation}
which is equal to the norm in Equation \eqref{eq:cepstralnorm}.

In this section, we will give this norm an interpretation in terms of poles and zeros of the underlying model, and the subspace angles of this model. Given a model $M$, these subspace angles are defined (see Equation \eqref{eq:anglesreformulated}) as
\begin{equation}
\left[M \sphericalangle M_\mathbb{1}\right] =\left[\Gamma_\infty\left(M\right) \sphericalangle \Gamma_\infty\left(M^{-1}\right)\right].
\end{equation}
This is equivalent to the angles between $M$ and a model, $M_\mathbb{1}$, with the identity function as transfer function. To interpret the norm, we need to consider three different cases:
\begin{itemize}
	\item minimum-phase, stable systems (i.e. poles and zeros of magnitude smaller than 1),
	\item maximum-phase, unstable systems (i.e. poles and zeros of magnitude greater than 1),
	\item the mixed case, with poles and zeros of magnitudes both smaller and greater than 1.
\end{itemize}
For simplicity, we always assume simple poles and zeros (i.e. multiplicites of 1) in the proofs presented here. Results carry over when this is not the case, but the derivations become much more involved.

This section starts with three Subsections, corresponding to the three different cases. The weighted cepstral distance is interpretable in terms of subspace angles in the minimum and maximum-phase case, but not in the mixed-phase case. A fourth Subsection will provide a link between the weighted cepstral distance from Equation \eqref{eq:cepstraldistance} and the norm \eqref{eq:cepstralnorm}. A method to assess phase-type and stability of a system based on data alone will be provided in Section \ref{sec:phasetype}.

\subsection{Minimum-phase stable interpretation}
\label{sub:minimumphase}

In the case of a discrete-time, completely minimum-phase and stable deterministic system, we write
\begin{equation}
Y_{min,s}(z) = \frac{b(z)}{a(z)}U_d(z),
\label{eq:minimumsystem}
\end{equation}
where $a(z)$ and $b(z)$ are now polynomials built from respectively stable poles and minimum-phase zeros of the system. This assures that both the model and its inverse are completely stable dynamical systems. All poles and zeros are furthermore assumed to be simple. The subscript in $U_d$ denotes the fact that we are dealing with deterministic inputs. ARMA-models fit into this case.
\subsubsection{Cepstral norm}
From Equation \eqref{eq:powercepstrumcoefficients}, we can readily see that in this case
\begin{equation}
c_h(k)=\sum_{j=1}^{p}\frac{\alpha_j^{|k|}}{|k|} - \sum_{j=1}^{q}\frac{\beta_j^{|k|}}{|k|} \hspace{17pt} \forall k\neq 0,
\label{eq:minimumphasecoefficients}
\end{equation}
and again $c_h(0) = g^\prime$. We then derive for the norm in Equation \eqref{eq:cepstralnorm},
\begin{equation}
\begin{aligned}
&||H_{min,s}||_c^2 = \sum_{k=1}^{\infty}k\left(c_{h}(k)\right)^2\\
&=\sum_{k=1}^{\infty}k\left(\sum_{j=1}^{p}\frac{\alpha_{j}^{k}}{k} - \sum_{j=1}^{q}\frac{\beta_{j}^{k}}{k}\right)^2\\
&=\log\frac{\displaystyle \prod_{i=1}^{p}\prod_{j=1}^{q}\left|1-\alpha_{i}\bar{\beta}_{j}\right|^2}
{\displaystyle
\prod_{i,j=1}^{p}\left(1-\alpha_{i}\bar{\alpha}_{j}\right)
\prod_{i,j=1}^{q}\left(1-\beta_{i}\bar{\beta}_{j}\right)},
\end{aligned}
\label{eq:minimumphasenorm}
\end{equation}
where the last equality holds true because of the series expansion 
\begin{equation}
\log(1-x) = -\sum_{k = 1}^{\infty}\frac{x^k}{k} \hspace{7pt} \forall |x|<1.
\label{eq:logseries}
\end{equation}

\subsubsection{Subspace angles}
Now, we show we can also interpret the subspace angles from Equation \eqref{eq:anglesreformulated} in terms of poles and zeros of the model. As discussed above Equation \eqref{eq:subspaceeigenvalues}, these angles are defined between the column spaces of two matrices. We can prove (see Appendix \ref{app:observability}) that the column space of the observability matrix of the system in Equation \eqref{eq:minimumsystem} can be expressed in terms of the (simple) poles, $\alpha_j$ of the system, as follows
\begin{equation}
\begin{aligned}
range(&\Gamma_j\left(M_{min,s}\right)) \\
&= range\left(\begin{matrix}
1 & 1 & \cdots & 1\\
\alpha_1 & \alpha_2 & \cdots & \alpha_{p} \\
\alpha_1^2 & \alpha_2^2 & \cdots & \alpha_{p}^2 \\
\vdots & \vdots & \ddots & \vdots \\
\alpha_1^{j-1} & \alpha_2^{j-1} & \cdots & \alpha_{p}^{j-1} \\							
\end{matrix}\right),
\end{aligned}
\label{eq:observabilityminimum}
\end{equation}
where $\Gamma_j\left(M_{min,s}\right)$ is the infinite observability matrix of the system in Equation \eqref{eq:minimumsystem} truncated at the $j$-th term.
Analogously, the column space of the inverse of the minimum-phase system can be expressed as
\begin{equation}
\begin{aligned}
range(&\Gamma_j\left(M_{min,s}^{-1}\right)) \\
&= range\left(\begin{matrix}
1 & 1 & \cdots & 1\\
\beta_1 & \beta_2 & \cdots & \beta_{q} \\
\beta_1^2 & \beta_2^2 & \cdots & \beta_{q}^2 \\
\vdots & \vdots & \ddots & \vdots \\
\beta_1^{j-1} & \beta_2^{j-1} & \cdots & \beta_{q}^{j-1} \\							
\end{matrix}\right),
\end{aligned}
\label{eq:observabilityinverseminimum}
\end{equation}
where $\Gamma_j\left(M_{min,s}^{-1}\right)$ is the infinite observability matrix of the inverse system.

Denote the complex conjugate transpose of a matrix with the superscript $^H$. We then introduce the notation
\begin{equation}
\begin{aligned}
P_{(1)(1)} &= \Gamma_j^H\left(M_{min,s}\right)\Gamma_j\left(M_{min,s}\right)\\
P_{(1)(-1)} &= \Gamma_j^H\left(M_{min,s}\right)\Gamma_j\left(M_{min,s}^{-1}\right)\\
P_{(-1)(1)} &= \Gamma_j^H\left(M_{min,s}^{-1}\right)\Gamma_j\left(M_{min,s}\right)\\
P_{(-1)(-1)} &= \Gamma_j^H\left(M_{min,s}^{-1}\right)\Gamma_j\left(M_{min,s}^{-1}\right).
\end{aligned}
\end{equation}
Now, starting from Equation \eqref{eq:subspaceeigenvalues}, with $\Gamma_j^H\left(M_{min,s}\right)$ and $\Gamma_j\left(M_{min,s}^{-1}\right)$ plugged in for $A$ and $B$ respectively, some algebra easily shows that
\begin{equation}
\begin{aligned}
&\log\prod_{i=1}^{n}\cos^2\theta_i \\
&=\lim_{j\to\infty}\log\det\left(P_{(1)(1)}^{-1}P_{(1)(-1)}P_{(-1)(-1)}^{-1}P_{(-1)(1)}\right)\\
&=\log\frac{\displaystyle
	\prod_{i,j=1}^{p}\left(1-\alpha_{i}\bar{\alpha}_{j}\right)
	\prod_{i,j=1}^{q}\left(1-\beta_{i}\bar{\beta}_{j}\right)}{\displaystyle \prod_{i=1}^{p}\prod_{j=1}^{q}\left|1-\alpha_{i}\bar{\beta}_{j}\right|^2}\\
&= -||H_{min,s}||_c^2,
\end{aligned}
\label{eq:equivalence}
\end{equation}
where the $\theta_i$ denote the subspace angles, $H_{min,s}$ is the transfer function $b(z)/a(z)$ from Equation \eqref{eq:minimumsystem} and the last equivalence follows from Equation \eqref{eq:minimumphasenorm}.

We have thus, for the minimum-phase, stable case, interpreted the cepstral norm in terms of subspace angles of the model, connecting them through the poles and zeros of the models. This shows that the cepstral norm is indeed an interpretable model norm.

\subsubsection{Stochastic inputs}
The stochastic case is a corollary of this result. De Cock and De Moor \citep{DeCockThesis,de2002subspace} define the cepstral norm of a signal $y$, generated by model $M$, with transfer function $H$, as 
\begin{equation}
||H||_c^2 = \sum_{k=1}^{\infty}k\left(c_{y}(k)\right)^2,
\end{equation}
(compared to our definition in terms of cepstra of the \emph{transfer functions} in Equation \eqref{eq:cepstraldistance}). In the stochastic case, however, the output cepstrum, $c_y$, and cepstrum of the transfer function, $c_h$, coincide. Indeed, the cepstrum coefficients of a white noise input, $c_{u_w}(k)$ fulfil the condition that
\begin{equation}
c_{u_w}(k) = 0, \hspace{0.5em} \forall k \neq 0,
\end{equation}
and they drop out of the expression for the cepstral distance. In this case we then have
\begin{equation}
c_y(k) = c_h(k), \hspace{0.5em} \forall k \neq 0,
\end{equation}
and the two definitions thus coincide.

\subsubsection{Autonomous systems}

In the case of an autonomous system (i.e. the sequence $u(k) = 0,\hspace{0.25em}\forall k$ in Equation \eqref{eq:timedomain}), the same observation as in the previous Subsection can be made, namely
\begin{equation}
c_y(k) = c_h(k), \hspace{0.5em} \forall k \neq 0,
\end{equation}
and the norm can be calculated from the output signal. Autonomous systems have only poles, but the rest of the interpretation remains the same, and we omit the details here. Unstable autonomous systems exist as well, but we will not discuss them in the next Subsection. The results of Subsection \ref{sub:maximumphase} also apply to this type of system, however.

\subsection{Maximum-phase unstable interpretation}

\label{sub:maximumphase}
In the case of discrete-time, completely maximum-phase and unstable dynamical systems, we assume that for the model
\begin{equation}
Y_{max,u}(z) = \frac{d(z)}{c(z)}U_d(z),
\label{eq:maxphasesystem}
\end{equation}
all zeros and poles that are the roots of polynomials $d(z)$ and $c(z)$, respectively, are unstable. All poles and zeros are assumed simple. Both the system and its inverse are completely unstable.

\subsubsection{Cepstral norm}
Again from Equation \eqref{eq:powercepstrumcoefficients}, we see that in this case
\begin{equation}
\begin{aligned}
c_h(k) = \sum_{j=1}^{s}\frac{\bar{\gamma}_j^{-|k|}}{|k|} - \sum_{j=1}^{r}\frac{\bar{\delta}_j^{-|k|}}{|k|} \hspace{17pt} \forall k\neq 0,
\label{eq:maximumphasecoefficients}
\end{aligned}
\end{equation}
and $c_h(0) = g^\prime$. The unstable poles and zeros thus appear as their inverses (see Appendix \ref{app:derivepowerspectrum} to see why this is so). Since these inverted poles and zeros are again of magnitude smaller than 1, we can proceed as in the minimum-phase, stable case. This leads to the following expression for Equation \eqref{eq:cepstralnorm}:
\begin{equation}
\begin{aligned}
&||H_{max,u}||_c^2 = \sum_{k=1}^{\infty}k\left(c_{h}(k)\right)^2\\
&=\sum_{k=1}^{\infty}k\left(\sum_{j=1}^{s}\frac{\gamma_{j}^{-k}}{k} - \sum_{j=1}^{r}\frac{\delta_{j}^{-k}}{k}\right)2\\
&=\log\frac{\displaystyle \prod_{i=1}^{s}\prod_{j=1}^{r}\left|1-\gamma^{-1}_{i}\bar{\delta}^{-1}_{j}\right|^2}
{\displaystyle
	\prod_{i,j=1}^{s}\left(1-\gamma^{-1}_{i}\bar{\gamma}^{-1}_{j}\right)
	\prod_{i,j=1}^{r}\left(1-\delta^{-1}_{i}\bar{\delta}^{-1}_{j}\right)},
\end{aligned}
\label{eq:maximumphasenorm}
\end{equation}
where $H_{max,u}$ is the transfer function $d(z)/c(z)$ from Equation \eqref{eq:maxphasesystem}.

\subsubsection{Subspace angles}
\begin{figure*}[]
	\begin{align}
	&||H_{mix}||^2_c \nonumber \\
	&=\sum_{k=1}^{\infty}k\left(\sum_{j=1}^{p}\frac{\alpha_{j}^{k}}{k} +\sum_{j=1}^{s}\frac{\gamma_{j}^{-k}}{k} - \sum_{j=1}^{q}\frac{\beta_{j}^{k}}{k} - \sum_{j=1}^{r}\frac{\delta_{j}^{-k}}{k}\right)^2 \label{eq:fullcepstrumnorm}\\
	&=\log\frac{\displaystyle\prod_{i = 1}^{p}\prod_{j = 1}^{q}|1-\alpha_i\bar{\beta}_j|^2\prod_{i = 1}^{p}\prod_{j = 1}^{r}|1-\alpha_i\bar{\delta}^{-1}_j|^2\prod_{i = 1}^{q}\prod_{j = 1}^{s}|1-\beta_i\bar{\gamma}^{-1}_j|^2\prod_{i = 1}^{s}\prod_{j = 1}^{r}|1-\gamma_i^{-1}\bar{\delta}^{-1}_j|^2}{\displaystyle\prod_{i,j = 1}^{p}(1-\alpha_i\bar{\alpha}_j)\prod_{i = 1}^{p}\prod_{j = 1}^{s}|1-\alpha_i\bar{\gamma}^{-1}_j|^2\prod_{i,j = 1}^{s}(1-\gamma^{-1}_i\bar{\gamma}^{-1}_j)\prod_{i,j = 1}^{q}(1-\beta_i\bar{\beta}_j)\prod_{i = 1}^{q}\prod_{j = 1}^{r}|1-\beta_i\bar{\delta}^{-1}_j|^2\prod_{i,j = 1}^{r}(1-\delta^{-1}_i\bar{\delta}^{-1}_j)} \nonumber
	\end{align}
	\vspace{5pt}
	\begin{center}
		\rule{0.8\textwidth}{0.4pt}
	\end{center}
\end{figure*}
Sending all the poles and zeros of the system to their inverses, amounts to an inversion of the plane with respect to the unit circle. It is a well-known result from inversive geometry that such a transformation is a conformal mapping of the plane, and thus preserves angles \citep[][p.479]{kay2011college}.
Therefore, computing the subspace angles in the inverted space will give the same results as computing them in the original space. However, in the inverted space, unstable/maximum-phase poles and zeros are mapped to their stable/minimum-phase inverses, which allows us to repeat the results from Section \ref{sub:minimumphase}.

We can thus avoid problems with the infinite limit of the observability matrices in Equation \eqref{eq:equivalence} if we invert the whole plane around the unit circle, which maps the (simple) poles and zeros from Equation \eqref{eq:maxphasesystem} to their inverses. We can then express the range of the observability matrix of the unstable, maximum-phase system $M_{max,u}$ as
\begin{equation}
\begin{aligned}
range(&\Gamma_j\left(M_{max,u}\right)) \\&= \left(\begin{matrix}
1 & 1 & \cdots & 1\\
1/\gamma_1 & 1/\gamma_2 & \cdots & 1/\gamma_{s} \\
1/\gamma_1^2 & 1/\gamma_2^2 & \cdots & 1/\gamma_{s}^2 \\
\vdots & \vdots & \ddots & \vdots \\
1/\gamma_1^{j-1} & 1/\gamma_2^{j-1} & \cdots & 1/\gamma_{s}^{j-1} \\							
\end{matrix}\right),
\end{aligned}
\label{eq:observabilitymaximum}
\end{equation}
with an analogous result in terms of (simple) zeros for the observability matrix of the inverse of the system, $\Gamma^{-1}_j\left(M_{max,u}\right))$.

Following the same procedure as in the previous Subsection, it is now straightforwardly shown that for the maximum-phase deterministic case
\begin{equation}
||H_{max,u}||_c^2 = -\log\left(\prod_{i = 1}^{n}\cos^2\theta_i\right),
\label{eq:maxphase}
\end{equation}
with $\theta_i$ the subspace angles of the model.

This shows that, for the maximum-phase, unstable case, the cepstral norm is again an interpretable model norm.

\subsection{Mixed-phase interpretation}
\label{sub:mixed}
Suppose now we have a combination of the previous two types of models, where there is a mixture of stable and unstable poles, and both minimum-phase and maximum-phase zeros. We can express such a system as
\begin{equation}
Y_{mix}(z) = \frac{b(z)d(z)}{a(z)c(z)}U_d(z),
\end{equation}
where $b(z)$ and $a(z)$ contain all stable zeros, $\beta_j$, and poles, $\alpha_j$, respectively, and $d(z)$ and $c(z)$ contain the unstable zeros, $\delta_j$, and poles $\gamma_j$. 

\subsubsection{Cepstral norm}
We now have the full expression for the cepstrum coefficients, as written down in Equation \eqref{eq:powercepstrumcoefficients}, which we repeat here:
\begin{equation}
\begin{aligned}
c_h(k) = &\sum_{j=1}^{p}\frac{\alpha_j^{|k|}}{|k|} + \sum_{j=1}^{s}\frac{\gamma_j^{-|k|}}{|k|} \\
- &\sum_{j=1}^{q}\frac{\beta_j^{|k|}}{|k|} - \sum_{j=1}^{r}\frac{\delta_j^{-|k|}}{|k|} \hspace{7pt} \forall k \neq 0,
\end{aligned}
\end{equation}
giving rise to the cepstrum norm in Equation \eqref{eq:fullcepstrumnorm}.

\subsubsection{Subspace angles}
\label{sub:mixedphasesubspace}
It is not clear how we can give the cepstral norm an interpretation in terms of subspace angles in this case. Directly calculating angles between rows of the observability matrix spanned by poles or zeros of magnitude larger than 1 and rows spanned by poles or zeros of magnitude smaller than 1, makes the limit in Equation \eqref{eq:equivalence} diverge, and renders us unable to prove the equivalence between the cepstral norm and the subspace angles between observability matrices.

Furthermore, Equation \eqref{eq:fullcepstrumnorm} shows us that there is no difference in norm between a system with only one stable pole $p$ and a system with only one unstable pole $1/p$, its inverse. Similarly, given any rational polynomial transfer function, we can invert any number of its stable or unstable poles or any minimum-phase or maximum-phase zeros and we will still get the same value for the norm. This is a direct consequence of the fact that the cepstral norm is based on the power spectrum, which is by definition insensitive to stability of poles and zeros, as can be readily seen from Equation \eqref{eq:powerspectrum}, where the transfer function is evaluated both in $z$ (the complex plane) and $z^{-1}$ (the complex plane inverted over the unit circle). This means that the norm for a stable, minimum-phase model and its unstable, maximum-phase equivalent generated by inverting all poles and zeros, would be equal (and, as we will see in the next subsection, the distance between them would $0$). For most applications where underlying dynamics are critical, this is not acceptable.

The interpretation of the model norm is thus lacking in this case, and we cannot give any meaningful system theoretical insight to the cepstral distance between signals coming from mixed-phase systems. The distance will therefore not be useful in this case. However, given a set of input-output signals, we do not a priori know the phase-type of the underlying model. In the Section \ref{sec:phasetype}, we will provide a data-driven way of assessing this phase-type (and thus the applicability of the cepstral distance), by employing the complex cepstrum.

\subsection{Relation between norm and distance}
\label{sub:normanddistance}

The cepstral norm in Equation \eqref{eq:cepstralnorm} can be related to the cepstral distance in Equation \eqref{eq:cepstraldistance}. We will show this, to simplify notation, for systems with only one stable pole and no zeros, but this result follows straightforwardly for all three cases considered in the previous subsections.

Given two signals, $y_1$ and $y_2$, coming from two systems, $M_1$ and $M_2$ with transfer functions
\begin{equation}
H_i(z) = \frac{1}{1-\alpha_iz^{-1}}, \hspace{0.5em}\forall i \in \{1,2\},
\end{equation}
with $|\alpha_i| < 1$. The power cepstra coefficients are then expressed, as in Equation \eqref{eq:powercepstrumcoefficients}, as
\begin{equation}
c_{h_i}(k) = \frac{\alpha_i^{|k|}}{|k|}, \hspace{0.5em}\forall k \neq 0, \hspace{0.5em}\forall i \in \{1,2\}.
\end{equation}
We now look at the cascade of transfer functions $H_{tot} = H_1H_2^{-1} = \frac{1-\alpha_2z^{-1}}{1-\alpha_1z^{-1}}$. Its power cepstra coefficients are
\begin{equation}
c_{h_{tot}}(k) = \frac{\alpha_1^{|k|}}{|k|} - \frac{\alpha_2^{|k|}}{|k|}, \hspace{0.5em}\forall k \neq 0
\end{equation}
which means its cepstral norm is
\begin{equation}
\begin{aligned}
||H_{tot}||_c^2 &= ||H_1H_2^{-1}||_c^2\\
&= \sum_{k=1}^{\infty}k\left(\frac{\alpha_1^{k}}{k} - \frac{\alpha_2^{k}}{k}\right)^2.\\
&= \sum_{k=1}^{\infty}k\left(c_{h_1}(k) - c_{h_2}(k)\right)^2 \\
&= d_c(y_1,y_2)^2.
\end{aligned}
\end{equation}

We have thus identified $d_c(y_1,y_2)^2 = ||H_1H_2^{-1}||_c^2$, which holds true in all cases considered in this paper. This means that, in going from the distance to the norm, zeros of $H_1$ and poles of $H_2$ will act as zeros of the cascaded system, $H_{tot}$ the norm and vice versa for the poles of $H_1$ and the zeros of $H_2$. All results obtained in this section could just as well be derived for the distance, but keeping track of poles and zeros of the different models results in very cumbersome notation.

Note that the interpretation of the cepstrum in terms of system matrices in Equation \eqref{eq:cepstrumsystemmatrices} allows us to not only calculate the distance between time series, but also between time series and state-space models. This can have applications in anomaly detection for industrial processes, where often a model for normal behaviour is known and a large distance between sensor signals and the model can indicate an anomaly.

\section{Assessing phase-type of systems}
\label{sec:phasetype}

Because of the use of the power spectral density in equation \eqref{eq:definitionpowercepstrum}, the power cepstrum does not take into account any information about phase-type and stability of the systems (i.e. whether poles and zeros have magnitudes greater or smaller than 1). Unsurprisingly, it is exactly this information that we need to discern between minimum-phase/stable, maximum-phase/unstable, or mixed models. This can be solved by instead employing the complex cepstrum from Equation \eqref{eq:complexcepstrumcoefficients}.

Start with a system \begin{equation}
Y(z) = \frac{b(z)d(z)}{a(z)c(z)}U_d(z),
\end{equation}
where $b(z)$ and $a(z)$ contain all stable zeros, $\beta_j$, and poles, $\alpha_j$, respectively, and $d(z)$ and $c(z)$ contain the unstable zeros, $\delta_j$, and poles $\gamma_j$. The degrees of these polynomials are unknown, and might be zero (i.e. no poles and zeros of that type are present in the system). The question we want to answer now, is whether we can, based on input-output signal pairs only, determine whether the system is mixed-phase or not. 

If the system is mixed-phase, the results of Subsection \ref{sub:mixedphasesubspace} apply, and we cannot straightforwardly interpret the norm in terms of subspace angles. If the system is either minimum-phase stable or maximum-phase unstable, the results of respectively Subsection \ref{sub:minimumphase} and \ref{sub:maximumphase} apply. The method in this Section thus serves as a test to determine whether the cepstral norm and the mathematical equivalences to subspace angles behind it apply to the input-output data at hand.

\begin{figure}[t]
	\centering
	\hspace{-2em}
	\includegraphics[width=\linewidth]{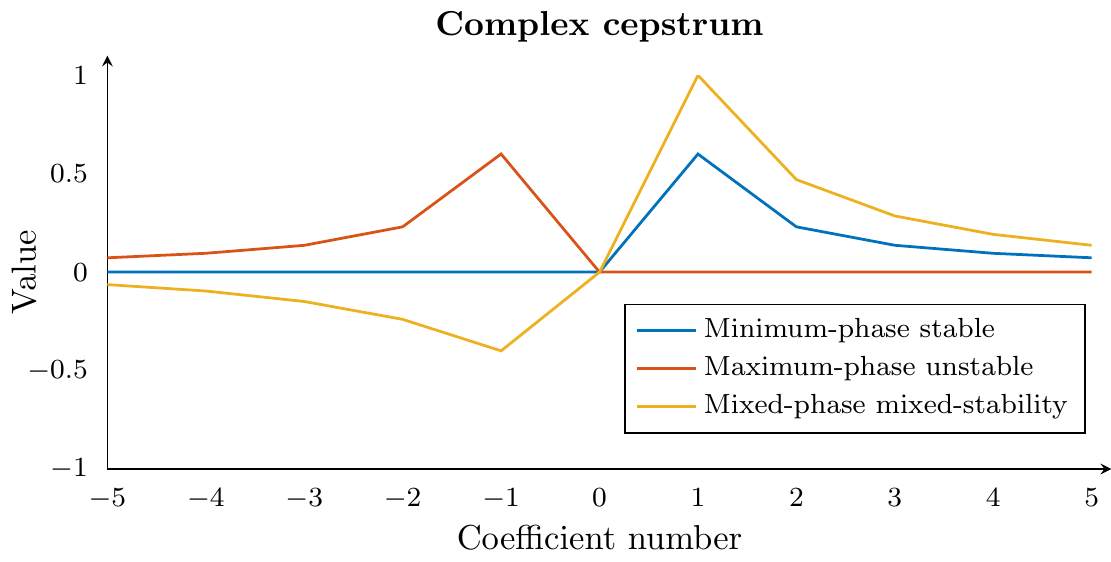} 
	\caption{The first five positive and negative complex cepstrum coefficients of respectively a minimum-phase stable system (zeros at $0.8$, $0.6$, $0$, poles at $0.9$, $0.7$, $0.4$, blue line), a maximum-phase unstable system (zeros at $1/0.8$, $1/0.6$, $\infty$, poles at $1/0.9$, $1/0.7$, $1/0.4$, red line) and a mixed-phase system (zeros at $1/0.8$, $0.6$, $0$, poles at $0.9$, $0.7$, $1/0.4$, yellow line). The zeroth cepstrum coefficient has been artificially set to 0 in all three cases to improve readability of the graph. This coefficient can take on large (positive or negative) values, but is not important for the discussion here.}
	\label{fig:complexcepstrum}                                
\end{figure}

We use the information contained in Equation \eqref{eq:complexcepstrumcoefficients} as a test to assess whether the cepstrum norm in Equation \eqref{eq:cepstralnorm} is interpretable in terms of subspace angles or not:
\begin{itemize}
	\item If the coefficients with negative coefficient number are all\footnote{As can be seen from equation \eqref{eq:complexcepstrumcoefficients}, the coefficients contain a factor $\frac{1}{k}$. This means that if the first few coefficients are zero, we can be sure that the following ones will also vanish. We thus do not have to test an infinite number of coefficients.} zero, but at least one of those with positive coefficient number is non-zero, the system is minimum-phase, and the results of Section \ref{sub:minimumphase} apply.
	\item If at least one of the coefficients with negative coefficient number is non-zero, but all of the ones with positive coefficient number are zero, the system is maximum-phase, and the results of Section \ref{sub:maximumphase} apply.
	\item If at least one of the coefficients with negative coefficient number is non-zero, and at least one of the ones with positive coefficient number is also non-zero, the system is mixed-phase, giving rise to the problems discussed in Subsection \ref{sub:mixedphasesubspace}.
\end{itemize}
Given an input-output signal dataset, we now have devised a technique that allows us to test whether the reformulated cepstrum distance makes sense. An example for each phase-type is shown in Figure \ref{fig:complexcepstrum}.

Note that this technique does not involve any unnecessary or extra calculations. Indeed, comparing the cepstrum coefficients in equation \eqref{eq:complexcepstrumcoefficients} with those in equations \eqref{eq:minimumphasecoefficients} and \eqref{eq:maximumphasecoefficients}, shows that the complex cepstra of the minimum-phase and maximum-phase case are nothing more than $\hat{c}_h(k),\hspace{4pt}\forall k>0$, and $\hat{c}_h(k),\hspace{4pt}\forall k<0$, respectively (up to a complex conjugate, which does not matter for the distance, as the complex conjugate pairs always appear together).

\begin{figure}[t]
	\centering
	\includegraphics[width=\linewidth]{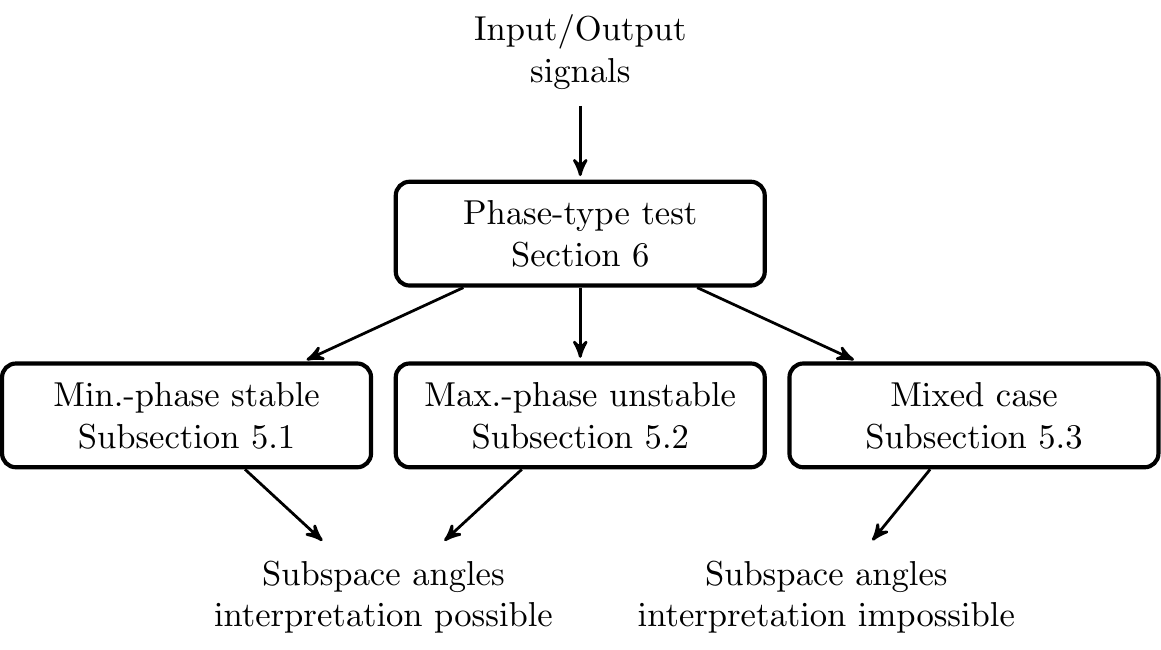} 
	\caption{A general overview for the procedure to interpret the norm in terms of subspace angles, starting from input/output signals. We first apply the test described in Section \ref{sec:phasetype}, to assess whether the system is minimum-phase/stable, maximum-phase/unstable or mixed (which is any case that is neither minimum-phase/stable or maximum-phase/unstable). Based on this classification, the results of respectively Subsections \ref{sub:minimumphase}, \ref{sub:maximumphase} or \ref{sub:mixed} apply. In the latter case, an interpretation in terms of subspace angles is impossible. In the other cases, this interpretation is possible.}
	\label{fig:generaloverview}                                
\end{figure}

The whole procedure of interpreting the cepstral norm in terms of subspace angles is summarized in Figure \ref{fig:generaloverview}.

\section{Numerical simulations}
\label{sec:numerical}

An iPython tutorial notebook is available on GitHub\footnote{https://github.com/Olauwers/Applicability-and-interpretation-of-the-deterministic-weighted-cepstral-distance}, containing a tutorial to check numerically the equivalences shown in this paper and tests the techniques presented for systems of different phase-types.

In time domain, we provide an implementation to check visually the phase of the underlying system (employing and verifying Equation \eqref{eq:complexcepstrumcoefficients}), then verify Equation \eqref{eq:equivalence} for stable, minimum-phase systems. For the unstable, maximum-phase case, we start from spectral data to visually check the phase of the underlying system (again employing and verifying Equation \eqref{eq:complexcepstrumcoefficients}) and proceed to verify Equation \eqref{eq:maxphase}. The reader is invited and encouraged to try out the tutorial for various systems, signal lengths, model orders, sampling times, \ldots

We test this for the minimum-phase/stable and maximum-phase/unstable systems in Figure \ref{fig:complexcepstrum}. Results are in good agreement with the theory. 
\begin{description}
	\item[Stable minimum-phase system] This system has zeros at $0.8$, $0.6$, $0$, and poles at $0.9$, $0.7$, $0.4$. We get an agreement between the weighted cepstral distance and the subspace angle distance (respectively the right hand side and left hand side of Equation \eqref{eq:equivalence}) of order $10^{-4}$. Computations are done starting from time domain input/output signal data. The agreement with the theoretical formulas in terms of poles and zeros, as in Equation \eqref{eq:minimumphasenorm}, is of order $10^{-4}$ for the weighted cepstral distance, and of order $10^{-13}$ for the subspace angle norm. Both norms approximate the expression in terms of poles and zeros, though the subspace angle norm is better at this.
	\item[Unstable maximum-phase case] This system has zeros at $1/0.8$, $1/0.6$, $1/10^{-15}$, and poles at $1/0.9$, $1/0.7$, $1/0.4$. Since in this case the system is unstable (and any output signals coming from it are unbounded), we start from data in the frequency domain. The agreement between the weighted cepstral distance and the subspace angle distance is again of order $10^{-4}$. The agreement with the theoretical formulas in terms of poles and zeros, as in Equation \eqref{eq:maximumphasenorm}, is of order $10^{-15}$ for the weighted cepstral distance, and of order $10^{-4}$ for the subspace angle norm. Both norms again approximate the expression in terms of poles and zeros, though this time, the weighted cepstral norm is better at this.
\end{description}

More examples can be found in the iPython notebook, and the reader is invited to test the equivalences in this paper on their systems/signals of choice.

\section{Conclusions and future work}
\label{sec:conclusions}

In this paper, we extended the subspace angles framework for the weighted cepstral distance from the stable, stochastic AR(MA) model case to its deterministic counterpart. We showed that reformulating the weighted distance in terms of the cepstrum of the transfer function, rather than the output cepstrum, leads to an extension that both reduces to the original stochastic case and readily generalizes to systems with deterministic inputs. We furthermore have proven that this reformulation also extends to unstable, maximum-phase systems, making use of the conformality of circle inversion.

The mixed-phase case, where both stable and unstable poles, and minimum-phase and maximum-phase zeros are present in the same model, remains an open problem. One can hope that a distance measure based on the complex cepstrum, thus taking into account phase information, would resolve this issue. Further research is needed, and will be pursued by the authors in a future paper.

For now, we have provided a way to assess phase-type and stability of the underlying dynamics of input/output signal pairs based on the raw data alone. This test not only allows us to assess whether the interpretation of the weighted cepstral distance in terms of subspace angles is valid for any given application, but also returns the cepstra of the input and output signals if it is valid. In this sense, this test is computationally `free', i.e. it does not introduce any extra computations.

We have formulated a data-driven distance measure, i.e. a method that needs nothing more than the input-output signal data, that provides us with insight into the underlying dynamical systems, allowing us to assess differences in dynamics, without ever having to identify the actual generating systems.

In future works, the results of this reformulation of the cepstrum distance will be applied to several engineering applications, such as signal clustering, classification and fault detection. An approximative on-line algorithm will also be implemented.

Finally, the extension to several more general model classes (i.e. MIMO, spatio-temporal, non-linear) remains an open problem.

\begin{ack}                               
O.L. is an SB PhD Fellow with the FWO. This research was supported in part by the Belgian Federal Science Policy Office: IUAP P7/19 (DYSCO, Dynamical systems, control and optimization, 2012-2017); 
the Flemish Government:
IWT: PhD grants - Industrial Research fund (IOF)
KU Leuven Internal Funds C16/15/059, C32/16/013;
imec strategic funding 2017.  
\end{ack}

\bibliographystyle{plain}        
\bibliography{DeterministicExtension}           

\appendix

\section{The cepstrum}
\label{app:computecepstrum}

An iPython notebook accompanying this paper, available on GitHub\footnote{https://github.com/Olauwers/Applicability-and-interpretation-of-the-deterministic-weighted-cepstral-distance}, implements the proposed techniques in this paper. In this appendix, we first give a derivation of the power cepstrum in terms of poles and zeros, and then expand on the computation of the cepstrum. We do this for an output signal $y$, but similar considerations hold for input signals and transfer functions.
\subsection{Deriving the power cepstrum}
\label{app:derivepowerspectrum}
In this subsection, we explicitly perform the analytic calculation in Equation \eqref{eq:definitionpowercepstrum}. A similar derivation can be found, for example, in \cite{oppenheim1975digital}.

Starting from equation \eqref{eq:poleszeros}, we can express the power spectrum $\Phi_h(z)$ as
\begin{equation}
\begin{aligned}
&\Phi_h(\textnormal{e}^{i\omega})\\ 
&= g\left|\frac{\prod_{j=1}^{q}\left(1 - \beta_j \textnormal{e}^{-i\omega}\right)\prod_{j=1}^{r}\left(1 - \delta_j \textnormal{e}^{-i\omega}\right)}{\prod_{j=1}^{p}\left(1 - \alpha_j \textnormal{e}^{-i\omega}\right)\prod_{j=1}^{s}\left(1 - \gamma_j \textnormal{e}^{-i\omega}\right)}\right|^2.
\end{aligned}
\end{equation}
Taking the logarithm, we write
\begin{equation}
\begin{aligned}
\log&(\Phi_h(\textnormal{e}^{i\omega})) = \log(g)\\
&+ \sum_{j=1}^{q}\left(\log\left(1 - \beta_j \textnormal{e}^{-i\omega}\right) + \log\left(1 - \overline{\beta}_j z\textnormal{e}^{i\omega}\right)\right)\\
&+ \sum_{j=1}^{r}\left(\log\left(1 - \delta_j \textnormal{e}^{-i\omega}\right) + \log\left(1 - \overline{\delta}_j \textnormal{e}^{i\omega}\right)\right)\\
&- \sum_{j=1}^{p}\left(\log\left(1 - \alpha_j \textnormal{e}^{-i\omega}\right) + \log\left(1 - \overline{\alpha}_j \textnormal{e}^{i\omega}\right)\right)\\
&- \sum_{j=1}^{s}\left(\log\left(1 - \gamma_j \textnormal{e}^{-i\omega}\right) + \log\left(1 - \overline{\gamma}_j \textnormal{e}^{i\omega}\right)\right).
\end{aligned}
\end{equation}
Using the series expansions from Equation \eqref{eq:logseries} and
\begin{equation}
\begin{aligned}
\log&(1-x\textnormal{e}^{-i\omega}) = -\log\left(\frac{x^{-1}\textnormal{e}^{i\omega}}{x^{-1}\textnormal{e}^{i\omega} - 1}\right)\\
&= -\log\left(-x^{-1}\textnormal{e}^{i\omega}\right) - \log\left(\frac{1}{1-x^{-1}\textnormal{e}^{i\omega}}\right)\\
&= -\log\left(-x^{-1}\textnormal{e}^{i\omega}\right) - \sum_{n = 1}^{\infty}\frac{x^{-n}}{n}\textnormal{e}^{in\omega} \hspace{7pt} \forall |x| > 1,\hspace{-0.5em} \label{eq:unstableexpansion}
\end{aligned}
\end{equation}
we then find
\begin{equation}
\begin{aligned}
\log&(\Phi_h(\textnormal{e}^{i\omega})) = \log(g)\\
&\hspace{-0.2em} - \sum_{j=1}^{q}\left(\sum_{n=1}^\infty \frac{\beta_j^n}{n}\textnormal{e}^{-in\omega} + \sum_{n=1}^\infty \frac{\overline{\beta}_j^n}{n}\textnormal{e}^{in\omega}\right)\\
&\begin{aligned}- \sum_{j=1}^{r}&\left(\log\left(-\overline{\delta}^{-1}\textnormal{e}^{-i\omega}\right)+\log\left(-\delta^{-1}\textnormal{e}^{i\omega}\right)\phantom{\sum_{n=1}^\infty \frac{\delta_j^{-n}}{n}\textnormal{e}^{in\omega}}\hspace{-2em}\right.\\&\hspace{0.65em}\left.+\sum_{n=1}^\infty \frac{\overline{\delta}_j^{-n}}{n}\textnormal{e}^{-in\omega} + \sum_{n=1}^\infty \frac{\delta_j^{-n}}{n}\textnormal{e}^{in\omega}\right)\end{aligned}\\
&\hspace{-0.2em}+ \sum_{j=1}^{p}\left(\sum_{n=1}^\infty \frac{\alpha_j^n}{n}\textnormal{e}^{-in\omega} + \sum_{n=1}^\infty \frac{\overline{\alpha}_j^n}{n}\textnormal{e}^{in\omega}\right)\\
&\begin{aligned}+ \sum_{j=1}^{s}&\left(\log\left(-\overline{\gamma}^{-1}\textnormal{e}^{-i\omega}\right)+\log\left(-\gamma^{-1}\textnormal{e}^{i\omega}\right)\phantom{\sum_{n=1}^\infty \frac{\gamma_j^{-n}}{n}\textnormal{e}^{in\omega}}\hspace{-2em}\right.\\&\hspace{0.35em}\left.+\sum_{n=1}^\infty \frac{\overline{\gamma}_j^{-n}}{n}\textnormal{e}^{-in\omega} + \sum_{n=1}^\infty \frac{\gamma_j^{-n}}{n}\textnormal{e}^{in\omega}\right)\end{aligned}\\
\end{aligned}\hspace{-2em}
\label{eq:logspectrum}
\end{equation}
Note now that $\log\left(-\overline{\delta}^{-1}\textnormal{e}^{-in\omega}\right)+\log\left(-\delta^{-1}\textnormal{e}^{in\omega}\right) = \log\left(\overline{\delta}^{-1}\textnormal{e}^{-in\omega}\delta^{-1}\textnormal{e}^{in\omega}\right) = -\log\left|\delta\right|^2$, and similarly for the terms $\log\left(-\overline{\gamma}^{-1}\textnormal{e}^{-i\omega}\right)+\log\left(-\gamma^{-1}\textnormal{e}^{i\omega}\right)$. We add these terms to the gain $g$, which now becomes $g^\prime$.

The power cepstrum coefficients $c_h(k)$ are now the inverse Fourier transform of $log\Phi_h$, or
\begin{equation}
\sum_{k=-\infty}^{\infty}c_y(k)e^{-ik\theta} = \log\Phi_y(e^{i\theta}).
\label{eq:inversefouriertransform}
\end{equation}
Comparing this with Equation \eqref{eq:logspectrum} (and noting that poles and zeros appear in complex conjugate pairs), we readily see that
\begin{equation}
\begin{aligned}
c_h(k) = &\sum_{j=1}^{p}\frac{\alpha_j^{|k|}}{|k|} + \sum_{j=1}^{s}\frac{\gamma_j^{-|k|}}{|k|} \\
- &\sum_{j=1}^{q}\frac{\beta_j^{|k|}}{|k|} - \sum_{j=1}^{r}\frac{\delta_j^{-|k|}}{|k|}
\end{aligned} \hspace{7pt} \forall k \neq 0,
\end{equation}
and
\begin{equation}
c_h(0) = g^\prime,
\end{equation}
which are the results in Equations \eqref{eq:powercepstrumcoefficients} and \eqref{eq:zerothcoefficient}.

\subsection{Computing the power cepstrum}
The computation of the power cepstrum is quite straightforward. Written down in pseudocode, the power cepstrum of a signal $y$ can be computed as
\begin{equation}
c_y(k) = \textnormal{IFFT}(\log(\Phi_y)),
\end{equation}
with $k = 1,\ldots,N$, and $N$ the length of the FFT. Both the implementation of the IFFT (\emph{Inverse Fast Fourier Transform}) \cite{brigham1988fast} and the logarithm are straightforward in this case, and are pre-implemented in systems theory-related packages in many commonly used scientific programming languages like MATLAB and Python.

Computing a good estimate of the power spectral density $\Phi_y$ is a little more involved. For long enough signals (from about $N = 2^{10}$ and beyond), we can employ the well-known \emph{Fast Fourier Transform}, or FFT, as an approximation of the Fourier transform and implement
\begin{equation}
\Phi_y(k) = \frac{1}{N}|\textnormal{FFT}(y)|^2,
\end{equation}
with $N$ again the length of the FFT and $|\cdot|$ denotes the magnitude. The FFT has a computational complexity of $\mathcal{O}(n\log{n})$.

Since the accuracy of the power spectral density estimate is very important to our techniques, it is, however, safer to use more accurate techniques, especially for shorter time series. One such algorithm is \emph{Welch's Method} \cite{welch1967use}.

Welch's Method divides the signal in overlapping windows, estimates the power spectral density of each of the windows using the FFT, and averages them out. The result is a less noisy estimate. This technique is again of $\mathcal{O}(n\log{n})$, and is what we will use most of the time.

For very short time series (less than $2^7$ time points), even more accurate estimation techniques exist (e.g. the \emph{Multitaper Method} \cite{percival1993spectral}), but these will be computationally more expensive, and are therefore to be avoided for longer signals. 

\subsection{Complex cepstrum}

Computing the complex cepstrum is a little more involved, due to the non-uniqueness of the complex logarithm. We will give a short overview of its computation here, a much more detailed account can be found in \citep[][Chapter 10]{oppenheim1975digital}. We will again assume an output signal $y$, with analogous results holding for input signals and transfer functions.

In principle, the complex cepstrum $\hat{c}_y(k)$ can be easily computed, starting from Equation \eqref{eq:definitioncomplexcepstrum}, in pseudocode, as
\begin{equation}
\hat{c}_y(k) = \textnormal{IFFT}(\log(\textnormal{FFT}(y))),
\end{equation}
with FFT and IFFT the Fast Fourier Transform and Inverse Fast Fourier Transform, respectively \cite{brigham1988fast}.

However, this expression is ambiguous. Evaluating the complex logarithm does not result in a unique solution. Indeed, the logarithm is the inverse of the exponential, and the complex exponential is a periodic function with period $2\pi$. Usually, this ambiguity is solved by demanding that $-\pi < \arg(Y(z)) < \pi$, which we will call the principal values of the phase, or the \emph{principal branch}. However, this results in samples of a discontinuous phase curve. For our purposes, we need samples of the continuous phase curve.

\begin{figure}[t]
	\centering
	\hspace{-2em}
	\includegraphics[width=\linewidth]{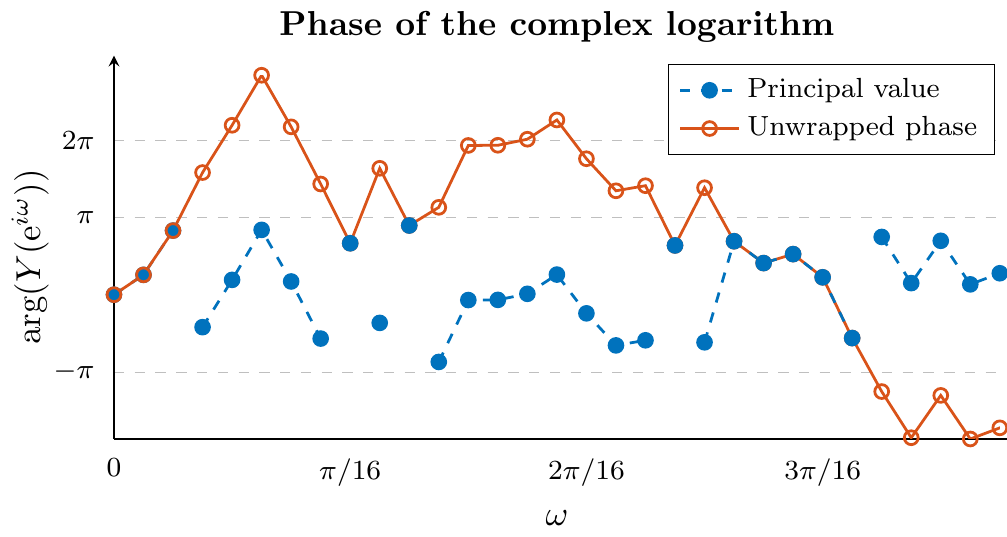} 
	\caption{Part of the phase of the Fourier transform of an output signal $y$. The principal values (shown in blue dots) are constrained to be between $-\pi$ and $\pi$. To compute the complex cepstrum, we \emph{unwrap} these principal values to obtain samples (shown by the red circles) from the continuous phase curve (an estimate of which is fitted through the samples with a red line). Where the dashed line between blue dots is interrupted, a multiple of $2\pi$ was added or subtracted to keep values between $-\pi$ and $\pi$.}
	\label{fig:phaseunwrap}                                
\end{figure}

We thus need to \emph{unwrap} the phase by adding or subtracting multiples of $2\pi$ to get samples from the continuous phase curve. Luckily, many methods exist to do this, and we use an off-the-shelf one (\emph{numpy.unwrap}) built into the Python NumPy library. Figure \ref{fig:phaseunwrap} shows the difference between the principal values of the phase and the unwrapped phase.

The complex cepstrum is then computed as in pseudocode Algorithm \ref{alg:complexcepstrum}.

\begin{algorithm}
\caption{Compute the complex cepstrum}
\label{alg:complexcepstrum}
\begin{algorithmic}[1]
	\Procedure{complex\_cepstrum}{y}
	\State $Y \gets FFT(y)$
	\State $\textnormal{angles} \gets \textnormal{unwrap}(\arg(Y)) $
	\State $Y^\prime \gets \log(|Y|) + i*\textnormal{angles}$
	\State $\hat{c}_y(k) \gets \textnormal{IFFT}(Y^\prime)$\\
	\Return $\hat{c}_y(k)$
	\EndProcedure
\end{algorithmic}
\end{algorithm}

\section{Observability matrices}
\label{app:observability}

\begin{figure*}[!h]
	\begin{equation}
	\begin{aligned}
	&\left[M^{(1)} \sphericalangle M^{(2)}\right]_r = \\
	&\lim_{j\to\infty}\left[\left(\left.Y_{0|j-1}^{(1)}\right|_{U_{0|j-1}^{(1)^\perp}} \left.U_{0|j-1}^{(2)}\right|_{Y_{0|j-1}^{(2)^\perp}}\right) \sphericalangle \left(\left.Y_{0|j-1}^{(2)}\right|_{U_{0|j-1}^{(2)^\perp}} \left.U_{0|j-1}^{(1)}\right|_{Y_{0|j-1}^{(1)^\perp}}\right)\right]
	\end{aligned}
	\label{eq:datadriven}
	\end{equation}
	\vspace{5pt}
	\begin{center}
		\rule{0.8\textwidth}{0.4pt}
	\end{center}
\end{figure*}

In this Appendix, we prove that the column space of the observability matrix of the system in Equation \eqref{eq:minimumsystem} can be expressed in terms of the poles, $\alpha_j$ of the system, as in Equation \eqref{eq:observabilityminimum}. We assume, for simplicity, that all poles have multiplicity one. Results carry over when this is not the case, but makes notation much more cumbersome. We then provide a computational way to find the range of the observability matrix.

\subsection{Range of the observability matrix in terms of poles}

Start with a stable, minimum-phase state-space model defined as in Equation \eqref{eq:timedomain}, with an observability matrix as in Equation \eqref{eq:observability}. We then diagonalize the matrix $A$ from the state-space model, and denote the diagonal matrix $\Lambda = TAT^{-1}$, with $T$ a suitable transformation matrix. The state space representation then becomes
\begin{equation}
\left\{
\begin{aligned}
x(k+1) &= \Lambda x(k) + B^\prime u(k)\\
y(k) &= C^\prime x(k) + D u(k)
\label{eq:timedomaintransformed}
\end{aligned}
\right.
,
\end{equation}
with $B^\prime = TB$ and $C^\prime = CT^{-1}$.

We can now write for the observability matrix 
\begin{align}
\Gamma_j &= \left(\begin{matrix}
C & CA & CA^2 & \cdots & CA^{j-1}
\end{matrix}\right)^\intercal \nonumber\\
&= \left(\begin{matrix}
	C^\prime & C^\prime\Lambda & C^\prime\Lambda^2 & \cdots & C^\prime\Lambda^{j-1}
\end{matrix}\right)^\intercal T \\
&= \left(\begin{matrix}
1 & 1 & \cdots & 1\\
\alpha_1 & \alpha_2 & \cdots & \alpha_{p} \\
\alpha_1^2 & \alpha_2^2 & \cdots & \alpha_{p}^2 \\
\vdots & \vdots & \ddots & \vdots \\
\alpha_1^{j-1} & \alpha_2^{j-1} & \cdots & \alpha_{p}^{j-1} \\							
\end{matrix}\right)
\left(
\begin{matrix}
c^\prime_1 & 0 & \cdots & 0 \\
0 & c^\prime_2 & \cdots & 0\\
0 & 0 & \ddots & 0\\
0 & 0 & \cdots & c^\prime_p
\end{matrix}\right)T, \nonumber
\end{align}
where the $c^\prime_i$ are the components of $C^\prime$ and the $\alpha_i$ are the elements of the diagonal matrix $\Lambda$, which coincide with the poles of the system.

As $C^\prime$ and $T$ are full-rank, we now know
\begin{equation}
\begin{aligned}
range(&\Gamma_j) \\
&= range\left(\begin{matrix}
1 & 1 & \cdots & 1\\
\alpha_1 & \alpha_2 & \cdots & \alpha_{p} \\
\alpha_1^2 & \alpha_2^2 & \cdots & \alpha_{p}^2 \\
\vdots & \vdots & \ddots & \vdots \\
\alpha_1^{j-1} & \alpha_2^{j-1} & \cdots & \alpha_{p}^{j-1} \\							
\end{matrix}\right),
\end{aligned}
\end{equation}
as in Equation \eqref{eq:observabilityminimum}, which is what we have set out to prove.

A similar reasoning leads us to the result in Equation \eqref{eq:observabilitymaximum} for the unstable, maximum-phase case.

\subsection{Computing the range of the observability matrix}
\label{app:computeangles}

An implementation of the techniques in this Appendix can be found in the accompanying iPython notebook.

Note that the formulation of subspace angles in terms of observability matrices in Equation \eqref{eq:anglesreformulated} allows for a data-driven way of calculating the subspace angles.

Start by denoting by $Y_{0|i-1}$ the $i\times j$ Hankel matrix composed of $i$ rows of elements of the time series $y$, starting at $y(0)$, where $j$ is the number of columns\footnote{Typically, for stationary time series, the limit case where $j \to \infty$ is considered.} considered ($i\ll j$), as follows:
\begin{align}
&Y_{0|i-1} = \nonumber\\
&\frac{1}{\sqrt{j}}
\begin{pmatrix}
y(0) & y(1) & y(2) & \cdots & y(j-1) \\
y(1) & y(2) & y(3) & \cdots & y(j) \\
y(2) & y(3) & y(4) & \cdots & y(j+1) \\
\vdots & \vdots & \vdots & \ddots & \vdots \\
y(i-1) & y(i) & y(i+1) & \cdots & y(i+j-2)
\end{pmatrix}.
\label{eq:hankelmatrix}
\end{align}
A similar structure can be defined for the input, and we will denote this Hankel matrix as $U_{0|i-1}$.

We can describe the relation between Hankel matrices (dropping subscripts for clarity) of inputs and outputs, as in equation \eqref{eq:hankelmatrix}, by the formula (see \cite{van2012subspace})
\begin{equation}
\begin{aligned}
Y &= \Gamma_jX + H_jU\\
\end{aligned}
\end{equation}
where the $X$'s denote the state sequence, $A$ is one of the system matrices of equation \eqref{eq:timedomain}, $\Gamma_j$ is the observability matrix up to the $j$-th term and $H_j$ and $\Delta_j$ are block matrices built from state-space system matrices that aren't too important for the present discussion. For an extensive discussion of the subspace framework, see \citep{van2012subspace}.

We can now easily see that
\begin{equation}
\left.Y\right|_{U^{\perp}} = \left.\Gamma_jX\right|_{U^{\perp}},
\end{equation}
where $Y|_{U^{\perp}}$ denotes the projection of the column space of $Y$ onto the orthogonal complement of the column space of $U$. We assume all of the matrices to be of full rank. It then follows from this equation that
\begin{equation}
range(\left.Y\right|_{U^{\perp}}) = range(\Gamma_j).
\end{equation}
The angles from Equation \eqref{eq:anglesreformulated} can then be written as in equation \eqref{eq:datadriven}, where the rows spanned by zeros are computed by using the inverse model (i.e. computing $\left.U\right|_{Y^{\perp}}$). This gives us a full data-driven approach to calculating the subspace angles between deterministic models.

Alternatively, starting from frequency-domain data, we can employ the \emph{FORSE}-algorithm from \cite{liu1996frequency}, which estimates the range of the observability matrix from frequency data samples.


\end{document}